 \documentclass{ws-ijmpa}

\usepackage[super,compress]{cite}
\usepackage{graphicx}
\usepackage{}
\begin{document}
\markboth{ROHIT TIWARI, JUHI OUDICHHYA and AJAY KUMAR RAI}{Mass-spectra of light-heavy tetraquarks}

%
\catchline{}{}{}{}{}
%

\title{Mass-spectra of light-heavy tetraquarks
\footnote{Spectroscopy of light-heavy flavoured tetraquark states.}
}

\author{ROHIT TIWARI, JUHI OUDICHHYA and AJAY KUMAR RAI}

\address{Department of Physics, Sardar Vallabhbhai National Institute of Technology\\
Surat, Gujarat 395007, INDIA
\\
rohittiwari843@gmail.com, oudichhyajuhi@gmail.com}

\author{ }


\maketitle

\begin{history}
\received{Day Month Year}
\revised{Day Month Year}
\end{history}

\begin{abstract}
The mass spectra of light-heavy tetraquarks $cq\bar{c}\bar{q}$ (q= u, d) are computed in a non-relativistic diquark model with one-gluon exchange plus confining potential. In the diquark model, a $cq\bar{c}\bar{q}$ state is regarded to be made of a light-heavy diquark (qc) and an antidiquark $\bar{q}\bar{c}$ in triplet and antitriplet colour configuration respectively. The masses of charm mesons were calculated in order to fit the model parameters used to create the masses of tetraquarks and therefore enhance the model's reliability. The masses of $(cq\bar{c}\bar{q})$ tetra-quark states are determined to be in the range of 3.8 GeV - 4.7 GeV, which is consistent with the experimentally reported charmonium-like states. In particular, the $Z_{c}(3900)$, $Z_{c}(4430)$, and $\psi(4660)$ tetraquarks, which have been seen experimentally, may all be described by our model. 
\keywords{Diquark; Exotic mesons; Tetraquarks.}
\end{abstract}

\ccode{PACS numbers:}


\section{Introduction}	

Since the early days of the quark model, the idea of exotic multiquark hadrons with valence quarks that are distinct from mesons $(q\bar{q})$ and baryons $(qqq)$ has been explored \cite{Y. R. Liu et al,S. L. Olsen}. The lack of compelling experimental evidences for such multiquark states has kept researchers at a distance. However, significant progress in experimental facilities (such as LHCb, Belle, CDF, D0, BESIII, BABAR, and CLEOc) of quarkonium studies has recently been made, resulting in the discovery of multiple exotic states X(3872), X(6900), Y(4260), $Z_c(3900)$, etc. \cite{X3872,cccc,Y4260,Z3900}. The simplest exotic multiquark system is a tetraquark, which consists of two quarks and two antiquarks. Heavy tetraquarks are of particular interest because the presence of a heavy quark increases the binding energy of the bound system and, as a result, the probability that such tetraquarks will have masses below the thresholds for decays to open heavy flavour mesons \cite{exotic1,exotic2,exotic3,exotic4}. Additionally, we emphasise the importance of the heavy-light tetraquarks $QQ\bar{q}\bar{q}$ (q=u,d) in a diquark-antidiquark configuration. It would be fascinating to investigate the existence of a $QQ\bar{q}\bar{q}$ tetraquarks that remains stable against strong decays, but there is no experimental proof yet \cite{QQqq1,QQqq2}. 
In 2003, the Belle collaboration \cite{X3872} discovered the exotic hadron X(3872) aka $\chi_{c1}(3872)$ in the decays $B^{\pm} \rightarrow K^{\pm} \pi^{+} \pi^{-} J/\Psi$. Following confirmation by BABAR \cite{X3872BABAR}, CDF II \cite{X3872CDFII}, D0 \cite{X3872D0}, and subsequently by the LHCb \cite{X3872LHCB} and CMS \cite{X3872CMS}, the X(3872) was established as a genuine resonance, as opposed to a threshold effect (labeled as "cusp"). It's discoveries proved to be a defining moment in hadron physics, ushering in a new era. Debris produced by B-meson decays led to the formation of tetraquarks (four quark states), each of which comprises one of the cc quark pairs \cite{Ahmed Ali}.
\\
\\
Long ago, Lipkin \cite{H. J. Lipkin} and Ader et al. \cite{J. P. Ader}, demonstrated that $QQ\bar{q}\bar{q}$ is theoretically stable against strong decays. Recently, it was shown that $bb\bar{q}\bar{q}$ is stable against strong decays, but not its charm counterpart, $cc\bar{q}\bar{q}$, or the combined (beauty charm) $bc\bar{q}\bar{q}$ state \cite{M. Karliner,E. J. Eichten}. See References \cite{A. Czarnecki,S. Q. Luo} for in depth discussions of the stability of various heavy-light tetraquarks. There has been no conclusive evidence for flavour exotic states in the lattice results, despite the fact that plausible $cu\bar{c}\bar{d}$ candidates with $J^{PC} = 1^{+-}$ have been investigated in the $D\bar{D}^{*}$ threshold region, whereas the $Z^{+}_{c}(3900)$ has been experimentally detected \cite{Z3900}. Despite mounting experimental evidence, no  $c\bar{c}s\bar{s}$ resonance has been discovered in $J/ \psi \phi$ scattering on the lattice, and in the $DD^{*}$ scattering, no $cc\bar{d}\bar{u}$ bound state has been discovered either \cite{sasa,exotic5}.
\\
\\
Theoretically, quarkonium physics is a well-developed field, with its origins in the non-relativistic quarkonium potential \cite{E. Eichten1978,E. Eichten1980}. It is currently expressed in terms of effective field theories based on QCD \cite{N. Brambilla,A. Gray} and lattice-QCD \cite{R.J. Dowdall,L. Liu}. The advancement of lattice-based approaches is remarkable, since it enables us to relate observable hadronic features to the basic parameters of quantum field theory \cite{Gyang,L. Leskovec}. The diquark model have been a successful tool to solve four body into two body problem, in Refs. \cite{prd,relativistic,Pan,d1,d2,d3,d4,d5,d6,d7,d8,d9,rohit} author's has succesfully obtained and predicted the the massses of tetraquarks in a diquark model.
\\ 
The study of hidden charm tetraquark states in a diquark model has been a successful tool to solve four body problem particularly, hadron spectra below the strong decay threshold are reliably estimated, and substantial multiplet findings are accessible.
The mass-spectra of light-heavy tetraquark in the hidden charm ($cq\bar{c}\bar{q}$) sectors are calculated in this paper. The mass-spectra of tetraquarks have been determined using a non-relativistic model by solving the $Schr\ddot{o}dinger$ equation. The article is structured as follows: Section II discusses the theoretical framework and formulation, while Section III discusses the results and discussion. Sec.IV contains the summary of this work.

\section{Theoretical framework}

The non-relativistic model with static potential is presented for the spectroscopic analysis of hadronic bound states containing heavy-heavy and heavy-light quarks in the diquark-antidiquark ($\mathcal{D-\bar{D}}$) approximation. We begin by converting the four-body system into three 2-body problems: a diquark [$QQ$] and antiquark [$\bar{Q}\bar{Q}$] system with two quarks and antiquarks pairs, respectively, and subsequently a diquark-antidiquark bound as the tetra-quark.
It was feasible to determine the mass spectra of the $cq\bar{c}\bar{q}$ tetraquark state using code originally developed by W. Lucha et al., by numerically solving the $Schr\ddot{o}dinger$ equation using the fourth-order Runge-Kutta (RK4) technique \cite{spin2}.
The optimal technique is to use the central potential \cite{prd} to solve two-body problems in the center-of-mass frame. The angular and radial terms of a wave function may be distinguished using spherical harmonics. The kinetic energy of quarkonium and tetraquarks may be calculated using the formula $\mu = \frac{M_{\mathcal{D}} M_{\mathcal{\bar{D}}}}{M_{\mathcal{D}}+ M_{\mathcal{\bar{D}}}}$, where $M_\mathcal{D}$ and $M_\mathcal{\bar{D}}$ are the masses of the constituents.\\
The kinetic energy of the heavy quark system is relatively smaller than the rest mass energy of the constituent quarks, thus employing static potentials in a non-relativistic model could be a viable approximation \cite{exotic2}. In the potential model, the spin-dependent terms are included perturbatively.  This approach generates a set of four free optimal parameters that can be used to generate the meson spectra, and later they can also be used to calculate the mass-spectra of diquarks and tetraquarks.
The Hamiltonian may be expressed in terms of an unperturbed one-gluon exchange (OGE) potential and a relativistic mass correction term $V^{1}(r)$ \cite{rohit}. The fundamental two-body hamiltonian center-of-mass frame of mesons and tetraquarks is denoted by the following:
\begin{equation}
H=\sum_{i=1}^{2}(M_{i}+\frac{p_{i}^{2}}{2M_{i}})+ V(r)
\end{equation}
Here $M_{i}$ is the constituent mass and $p_{i}$ is the relative momentum of the system, while $ V(r)$ is the interaction potential.
\\
The time-independent radial $Schr\ddot{o}$dinger \cite{scr1,scr2} equation for two body problem can be expressed as; 

\begin{eqnarray}
\left[-\frac{1}{2\mu}\left(\frac{d^{2}}{d(r)^{2}}+\frac{2}{r}\frac{d}{d(r)}-\frac{L(L+1)}{r}^{2}\right)+ V(r)\right] \psi(r) = E\psi(r) \quad \quad  
\end{eqnarray}

where, L and E are the orbital quantum number and energy eigenvalue respectively. By substituting $\psi(r) = r^{-1}\phi(r)$ in Eq.(1) modifies to;
\begin{equation}
\left[-\frac{1}{2\mu}\left(\frac{d^{2}}{dr^{2}}+\frac{L(L+1)}{r^{2}}\right)+V(r)\right]  \phi(r) = E\phi(r)  
\end{equation}
A zeroth-order $V(r)$ cornell-like potential \cite{cornell} is a reliable and extensively used potential model in the spectroscopic analysis of heavy-quarkonium systems. The Cornell-like potential $V_{C+L}(r)$ is composed of the Coulomb and linear term, with the coulomb part arising from a Lorentz vector exchange (basically one gluon exchange) and the linear term causing confinement typically associated with a Lorentz scalar exchange.
\begin{eqnarray}
V_{C+L}(r)=\frac{k_{s}\alpha_{s}}{r}+br\\
k_{s} = -\frac{4}{3} \quad for \quad q\bar{q}\\
= -\frac{2}{3} \quad for \quad qq \quad or \quad \bar{q}\bar{q}
\end{eqnarray}
where, $\alpha_{s}$ is known as the QCD running coupling constant, $k_ {s}$ is a colour factor, b is string tension. We have included the relativistic mass correction term $V^{1}(r)$ originally established by Y. koma et al. \cite{koma}, in the central potential. The final form of central potential is provided by;
\begin{equation}
 V(r) = V_{C+L}(r) + V^{1}(r)\left( \frac{1}{M_\mathcal{D}}+\frac{1}{M_\mathcal{\bar{D}}}\right)+\mathcal{O}\left(\frac{1}{m^{2}}\right)
\end{equation}
$M_\mathcal{D}$ and $M_\mathcal{\bar{D}}$ are the masses of diquark and an antidiqaurk respectively.
The non-perturbative form of relativistic mass correction term $V^{1}(r)$ is not yet known, but leading order perturbation theory yields \cite{koma},
\begin{equation}
V^{1}(r)=-\frac{C_{F}C_{A}}{4} \frac{\alpha^{2}_{s}}{(r)^{2}}
\end{equation}
where $C_{F}=\frac{4}{3}$ and $C_{A}=3$  are the Casimir charges of the fundamental and the adjoint representation respectively \cite{koma}. When applied to charmonium, the relativistic mass correction is found to be equivalent to the coulombic component of the static potential, and one-fourth of the coulombic term for bottomonium \cite{ak,vk,dpepjc,dpepjp,dpijp1,dpfew,dpijp2}. 
\\
We introduce three spin-dependent interactions (spin-spin $V_{SS}(r)$, spin-orbit $V_{LS}(r)$, and tensor $V_{T}(r)$) for one gluon exchange based on the Breit-Fermi Hamiltonian, which will be solved using first order perturbation theory by adding their matrix components as corrections to the energy \cite{spin1,spin2,rohit}.
\begin{equation}
V_{SD} (r) = V_{SS} (r)+V_{LS} (r)+V_{T} (r),
\end{equation}
These spin-dependent terms of Eq. (9) may be expressed as in terms of the vector and scalar sections of the static potential V(r). These spin-dependent interactions are included perturbatively to the potential and it needs a laborious algebra, which is not discussed in depth here, rather one can get the detailed discussion in the following Refs. \cite{debastiani,griffiths,prd,bethe,thesis}.
\begin{equation}
V_{SS} (r_{ij}) = C_{SS}(r_{ij})  S_{1} \cdot S_{2},
\end{equation}
\begin{equation}
V_{LS} (r_{ij}) = C_{LS}(r_{ij}) L \cdot S,
\end{equation}
\begin{equation}
V_{T} (r_{ij}) = C_{T}(r_{ij}) S_{12},
\end{equation}

Because the tetraquark radial wavefunction is obtained by treating the diquark and antidiquark as two body problem, it is reasonable to assume that the radial-dependence of the tensor term is the same for these four [$q\bar{q}$] interactions and can be obtained using the radial wavefunction. The following functional form for spin $\frac{1}{2}$ particles does not use any specific relation or eigenvalues, instead relying on general angular momentum elementary theory \cite{tb}. Within this approximation, generalization of tensor operator can be consider a sum of four tensor interaction between four quark-antiquark pair as illustrated in \cite{thesis}. A thorough discussion on tensor interaction can be found in Ref. \cite{thesis}.
In the present work there are four fitting parameters (m, $\alpha_{s}$, b, $\sigma$) for which the model mass ($M_{i}^{f}$) of the particular tetraquark states have been calculated. 
\begin{center}
0.05 $\leq$ $\alpha_{s}$ $\leq$ 0.70\\
 0.01 Ge$V^2$ $\leq$ b $\leq$ 0.40 Ge$V^2$
 \\
 0.05 GeV $\leq$ $\sigma$ $\leq$ 1.50 GeV
 \\
 1.00 GeV $\leq$ $m_{c}$ $\leq$ 2.00 GeV
 \\
 0.3 GeV $\leq$ $m_{q}$ $\leq$ 0.350 GeV\\
\end{center}
The quark masses  $m_{c}$ = 1.4 GeV, $m_{q}$ = 0.330 GeV have been taken from PDG \cite{pdg}. From the above range, the fitted parameters are tabulated in Table:1 to obtain the mass-spectra of mesons, diquarks and tetraquarks.  The mass-spectra of charmonium mesons ($c\bar{q}$) and charm-light (anti)diquark have been obtained from data set I.  The mass-spectra of tetraquarks are computed using data set II.
\begin{table}[h]
\label{ta1}
\tbl{Fitting parameters}{
\begin{tabular}{cccc} \toprule
Data Set & $\alpha_{s}$ & $\sigma$(GeV) & b (Ge$V^{2}$) \\ \colrule
I\hphantom{000} & \hphantom{0}0.70 & \hphantom{0}1.10 & 0.10 \\
II\hphantom{000} & \hphantom{0}0.5167 & \hphantom{0}0.7045 & 0.1023 \\
 \botrule
\end{tabular}
} 
\end{table}

\section{Results and Discussion}
\subsection{Masses of D-mesons}
\label{sec:2}
To calculate the mass-spectra of diquarks and tetra-quarks, first, we estimate the mass-spectra of quarkonium states [$c\bar{q}$] whose results are tabulated in Table:1. 
The SU(3) color symmetry allows only colorless quark combination $|Q\bar{Q}\rangle$ to form any color  singlet state \cite{griffiths,thesis}, as in our case [$c\bar{q}$] is meson and exhibits $|Q\bar{Q}\rangle: \mathbf{3 \otimes\bar{3}=1\oplus 8}$ representation which leads to carry a color factor $k_{s}=-\frac{4}{3}$ \cite{debastiani}. The masses of the particular [$c\bar{q}$] states are obtained namely
\begin{equation} 
M_{(c\bar{q})} = M_{c}+ M_{\bar{q}}+ E_{c\bar{q}} + \langle V^{1}(r)\rangle
\end{equation}
The final masses obtained from the above expression constitute the contributions from different spin-dependent terms (spin-spin, spin-orbital and tensor) have tabulated in Table:1. The mass-spectra of the mesons produced in this study are compatible with the experimental data available in the most recent updated PDG \cite{pdg}.


\newpage
\label{tab:2}
Table 2: The Mass-Spectra of D-mesons [$c\bar{q}$], generated from data set I.
\rotatebox{90}{
\small
\begin{tabular*}{180mm}{@{\extracolsep{\fill}}ccccccccccccc}
\hline
$N^{2S+1}L_{J}$	& $J^{PC}$&	$\langle E\rangle$	&	$\langle V_{V}\rangle$	&	$\langle V_{S}\rangle$	&	$\langle V_{SS}\rangle$	&	$\langle V_{LS}\rangle$	&	$\langle V_{T}\rangle$ &	$\langle V^{(1)}(r) \rangle $	&	$\langle K.E. \rangle$	&	$M_{f}$  & $M_{Exp}$& Meson \\
\hline
$1^{1}S_{0}$&$0^{-+}$	&	269.1	&	-403	&	318	&	-127.2	&	0	&	0	&	-8.2 	&	481	&	1872	&	 1869&$D^{\pm}$\\
$1^{3}S_{1}$&$1^{--}$	&	269.8	&	-403	&	318	&	42.4	&	0	&	0	&	-7.5	&	310	&	2040	&	2010&$D^{*}(2010)^{\pm}$\\
$2^{1}S_{0}$&$0^{-+}$	&	859.1	&	-232	&	655	&	-73.1	&	0	&	0	&	-4.3	&	510	&	2515	&2549$\pm$19&$D_{0}(2550)^{0}$\\
$2^{3}S_{1}$&$1^{--}$	&	859.1	&	-232	&	655	&	24.5	&	0	&	0	&	-4.5	&	415	&	2616	&	2637$\pm$6& $D^{*}(2640)^{\pm \#\#}$ \\
$3^{1}S_{0}$&$0^{-+}$	&	1288.5	&	-179	&	921	&	-58.1	&	0	&	0	&	-3.1	&	604&	2960	&	-&	-\\
$3^{3}S_{1}$&$1^{--}$	&	1288.5	&	-179	&	921	&	19.7	&	0	&	0	&	-3.1	&	528	&	3039	&-&	-\\
$4^{1}S_{0}$&$0^{-+}$	&	1651.1	&	-151	&	1153&	-49.8	&	0	&	0	&	-2.3	&	699	&	3331	&	-&	-\\
$4^{3}S_{1}$	&$1^{--}$&	1651.1	&	-151	&	1153	&	16.6	&	0	&	0	&	-1.9	&	633	&	3399	&	-&	-\\
%

$1^{3}P_{0}$&$0^{++}$	&	699.3	&	-202	&	535	&	1.5	&	-77.4	&	-41.4	&	-5.5	&	365	&	2312	&	2343$\pm$10&$D_{0}^{*}(2300)^{0}$\\

$1^{3}P_{1}$&$1^{++}$	&	695.8	&	-202	&	535	&	1.5	&	-38.4	&	20.7	&	-5.5	&	362	&	2409	&2421& $D_{1}(2420)^{0}$\\

$1^{1}P_{1}$&$1^{+-}$	&	695.0	&	-202	&	535	&	-4.5	&	0	&	0	&	-5.4	&	368	&	2421	&	2422& $D_{1}(2420)$\\

$1^{3}P_{2}$&$2^{++}$	&	695.0	&	-202	&	535	&	1.5	&	38.4	&	-4.1	&	-5.5	&	362	&	2462	&	2461&$D_{2}^{*}(2460)$\\

$2^{3}P_{0}$&$0^{++}$	&	1135.4	&	-155	&	816	&	1.9	&	-77.8	&	-37.9	&	-3.5	&	472	&	2751	&-&	-	\\
$2^{3}P_{1}$&$1^{++}$	&	1143.5	&	-155	&	816	&	1.9	&	-38.4	&	18.5	&	-3.4	&	480	&	2855	&	-&	-\\
$2^{1}P_{1}$&$1^{+-}$	&	1143.0	&	-155	&	816	&	-5.9	&	0	&	0	&	-3.5	&	488	&	2867	&	-&	-\\
$2^{3}P_{2}$	&$2^{++}$ &	1143.0	&	-155	&	816	&	1.9	&	39.0	&	-3.7	&	-3.5	&	480	&	2910	&-&	-\\
$3^{3}P_{0}$	&$0^{++}$&	1517.3	&	-130	&	1057	&	2.2	&	-78.4	&	-36.2	&	-2.5	&	588	&	3135	&	-&	-\\
$3^{3}P_{1}$	&$1^{++}$&	1518.3	&	-130	&	1057	&	2.3	&	-39.4	&	18.1	&	-2.6	&	589	&	3229	&	-&	-\\
$3^{1}P_{1}$&$1^{+-}$	&	1518.4	&	-130	&	1057	&	-6.7	&	0	&	0	&	-2.6	&	598	&	3241	&	-&	-\\
$3^{3}P_{2}$&$2^{++}$	&	1517.3	&	-130	&	1057	&	2.3	&	39.2	&	-3.6	&	-2.5	&	589	&	3286	&	-&	-\\
$1^{3}D_{1}$&$1^{--}$	&	829.4	&	-164	&	609	&	0.08	&	-11.2	&	-5.1	&	-3.8	&	383	&	2762&2781$\pm$22	&$D^{*}_{1}(2760)^{0}$\\

$1^{3}D_{2}$&$2^{--}$	&	828.6	&	-164	&	609	&	0.08	&	-3.7	&	5.1	&	-3.6	&	383	&	2780	& -&-\\
$1^{1}D_{2}$&$2^{-+}$	&	829.5	&	-164	&	609	&	-0.2	&	0	&	0	&	-3.7	&	383	&	2778	&2747$\pm$6 & $D_{2}(2740)^{0}$\\

$1^{3}D_{3}$&$3^{--}$	&	828.8	&	-164	&	609.4	&	0.08	&	7.5	&	-1.4	&	-3.7	&	383	&	2785	&	2763$\pm$2.3 &$D^{*}_{3}(2750)$\\
$2^{3}D_{1}$	&$1^{--}$&	1175.8	&	-136	&	831	&	0.1	&	-15.1	&	-4.9	&	-2.6	&	480	&	3105	&	-&	-\\
$2^{3}D_{2}$&$2^{--}$	&	1176.4	&	-136	&	831	&	0.1	&	-5.1	&	4.9	&	-2.0	&	481	&	3127	&	-&	-\\
$2^{1}D_{2}$&$2^{-+}$	&	1177.5	&	-136	&	831	&	-0.4	&	0	&	0	&	-2.4	&	482	&	3127	&	-&	-\\
$2^{3}D_{3}$&$3^{--}$	&	1242.3	&	-136	&	831	&	0.1	&	10.1	&	-1.4	&	-2.5	&	481	&	3136	&	-&	-\\
[1ex]
\hline
\end{tabular*}
}

{\footnotesize $^{\#\#}$ The quantum numbers of these mesons are not assigned yet in the Recent updated PDG.}\\
There are a total of 11 charmed mesons ($c\bar{q}$) produced from the model, all of which have masses fairly closed to those predicted experimentally. Additionally, the current work's findings are consistent with those in Ref. \cite{prd}, where the author computed the mass-spectra of heavy-light tetraquarks [$Qq\bar{Q}\bar{q}$] (Q = b, c and q = u, d) including all heavy tetraquarks.
 In the case of S-wave charmed mesons states the discrepancy is around 30 MeV. Particularly, in second radial states where the strength of spin-spin interaction declines sharply which leads to maximizes the discrepancy.
At high energy scale, discrepancy nearly 30-60 MeV's between the model's mass and experimental data can be tolerated and the fitting parameters are assumed as best fit. 


\subsubsection{Diquarks}
A (anti)diquark ($\mathcal{(\bar{D})D}$)is a pair of (anti)quarks that interact with one another through gluonic exchange and can form a bound state \cite{exotic4}.
The Pauli principle should also be considered, which results in the following ground state diquark limitations.
 The (qq') diquark, which is made of quarks of various flavours, may have spins S = 0,1 (scalar [qq'], axial vector $\lbrace qq'\rbrace$ diquarks), while the $\lbrace qq\rbrace$ diquark, which is composed of quarks of the same flavour, can only have spin S = 1. Because of the stronger attraction owing to the spin–spin interaction, the scalar S diquark is frequently referred to as a ``good" diquark, while the heavier axial vector diquark is referred to as a ``bad" diquark \cite{exotic2}. To produce the most compact diquark, we will utilise the ground state ($1^{1}S_{0}$) diquarks [$cq$], which have no orbital or radial excitations.
\\
%
According to QCD color symmetry, two quarks are combined in the fundamental ({\bf3}) representation to obtain the diquark, presented by  $\mathbf{3 \otimes 3 = \bar{3} \oplus 6}$.
Moreover, antiquarks are combined in the $\mathbf{\bar{3}}$ representation and can be presented as $\mathbf{\bar{3} \otimes \bar{3} = 3 \oplus \bar{6}}$ \cite{debastiani,griffiths}. 
The diquark-antidiquark approximation is significant because it reduces a complex four-body problem to a simple two-body problem. The hamiltonian, on the other hand, ceases replicating the meson spectra when doing the full four-body  basis treatment \cite{diquark1}. The explanation for this is simple: the $3 \otimes \bar{3}$ color coupling can be transformed into a $1\otimes1$ state, and also a $8\otimes8$ state. The QCD color symmetry  produces a color factor $k_{s}=-\frac{2}{3}$ in antitriplet state and makes the short distance part $(\frac{1}{r})$ of the interaction attractive \cite{griffiths}.
We compared the diquark masses acquired in this work to those obtained in the other prior investigations mentioned in Table:3.
\setcounter{table}{2}
\begin{table}
\label{tab:4} 
\tbl{The masses vector (anti)diquarks  and energy eigenvalue from present work and comparison with other prior works. Units are in (MeV).}{     
\begin{tabular}{ccccccccc}
\hline
Diquark & $\langle E^{0}\rangle$ & Ours &\cite{exotic4} & \cite{prd} & \cite{de}& \cite{LM} &\cite{RK}&\cite{mh}\\
\hline
 cq &258.3 & 1963 &1973& 2018 & 2036 & 1933  & 1865&1973\\
\noalign{\smallskip}\hline
\end{tabular}
}
\end{table}
The masses of diquarks calculated in this research are consistent with \cite{exotic4,LM} and are less than those reported in Ref. \cite{prd,de}. The discrepancies may be due to the addition of new and updated data in this study.

\subsection{Masses of Tetraquarks}
Tetraquarks are color singlet states made up of a diquark $(\mathcal{D})$ and an antidiquark ($\mathcal{\bar{D}}$) in color antitriplet $\mathbf{\bar{3}}$ and triplet $\bf{3}$ configurations respectively, that are held together by color forces \cite{debastiani,prd,de}. 
A $T_{Qq\bar{Q}\bar{q}}$ is color singlet states and yield a color factor $k_{s} = -\frac{4}{3}$.
The ($1^{1}S_{0}$) diquark(antidiquark) are combined to form color singlet tetraquark \cite{thesis}, and that can be represented as; $|QQ|^{3} \otimes|\bar{Q}\bar{Q}|^{\bar{3}}\rangle = \mathbf{1  \oplus 8}$. The mass-spectra of heavy-light tetraquarks ($cq\bar{c}\bar{q}$) have been obtained with the same formulation as in the case of mesons, namely;
\begin{equation}
M_{cq\bar{c}\bar{q}} = M_{cq}+ M_{\bar{c}\bar{q}} + E_{[cq][\bar{c}\bar{q}]} + \langle V^{1}(r)\rangle
\end{equation}
All spin-dependent terms have been computed for spin-1 diquarks and antiquarks that combine to produce a color singlet tetraquark with spin $S_{T}$ = 0,1,2. 
The interaction of $S_{T}$ with the orbital angular momentum $L_{T}$ results in the formation of a color singlet state $S_{T} \otimes L_{T}$.
\begin{equation}
|T_{4Q}\rangle = |S_{\mathcal{D}},S_{\mathcal{\bar{D}}},S_{T},L_{T} \rangle_{J_{T}}
\end{equation}
To find out the quantum numbers ($J^{PC}$) of the tetra-quark states, one can use the following formula; $P_{T}=(-1)^{L_{T}}$ and $C_{T}=(-1)^{L_{T} + S_{T}}$.
The masses of low-lying S-wave $cq\bar{c}\bar{q}$ states are anticipated to be in the range of 3.8-4.5 GeV \cite{exotic4,de,mh}, in the current study as well the masses are also found to be in this range.As shown in Table:4, the compactness of the 1S-wave states are mostly due to the coulomb interaction. This indicates that one-gluon exchange is the dominant mechanism behind the strong interaction between diquarks and antidiquarks, which results in a negative energy eigenvalue E. The contribution of the confinement term increases with the increase in radial and orbital states.
\\
Within the specific tetraquark mass-spectrum, the attractive strength of the spin-spin interaction decreases as the number of radial and orbital excited states increases. In this instance, we must bear in mind that the factors originating from $S_{1}$ and $S_{2}$ are greater for the coupling of two spin-1 particles than for the coupling of two spin-$\frac{1}{2}$ particles. It is worth noting that, despite the fact that the spin-dependent terms have been suppressed by a factor $\frac{1}{m_{qq}^{2}}$, one would anticipate them to be less than the equivalent terms in $q\bar{q}$ mesons. The color interaction brings diquark and antidiquark so close together that the suppression caused by this component $\frac{1}{m_{qq}^{2}}$, is swamped by the massive suppression at the system's origin. It is possible to state that spin-dependent factors in $\mathcal{D\bar{D}}$ interactions lead to a minor contribution to the masses of tetraquarks. 
\newpage
 Table 4: The Mass-Spectra of [$cq\bar{c}\bar{q}$] tetraquark, generated from data set II.

\rotatebox{90}{
\small
\begin{tabular*}{180mm}{@{\extracolsep{\fill}}cccccccccccccc}
\hline
 $N^{2S+1}L_{J}$&$J^{PC}$	&	$\langle E^{0}\rangle$	&	$\langle V^{(0)}_{V}\rangle$	&	$\langle V^{(0)}_{S}\rangle$	&	$\langle V^{(0)}_{SS}\rangle$	&	$\langle V^{(1)}_{LS}\rangle$	&	$\langle V^{(1)}_{T}\rangle$ &	$\langle V^{(1)}(r) \rangle $	&	$\langle K.E. \rangle$	&	$M^{i}_{f}$  & $M_{th}$&Threshold\\
\hline
$1^{1}S_{0}$	&	$0^{++}$	&	-45.6	&	-592	&	169	&	-69.1	&	0	&	0	&	-3.7	&	446	&	3811&3738&$D^{\pm}D^{\pm}$\\
$1^{3}S_{1}$	&	$1^{+-}$	&	-45.6	&	-592	&	169	&	-34.5	&	0	&	0	&	-3.7	&	412	&	3846 &3879&$D^{\pm} D^{*\pm}$\\
$1^{5}S_{2}$	&	$2^{++}$	&	-45.6	&	-592	&	169	&	34.5	&	0	&	0	&	-3.7	&	342	&	3914& 4020 &$D^{*\pm} D^{*\pm}$ \\
$2^{1}S_{0}$	&	$0^{++}$	&	447.3	&	-290	&	397	&	-19.4	&	0	&	0	&	-1.7	&	359	&	4353& ...& ...\\
$2^{3}S_{1}$	&	$1^{+-}$	&	449.4	&	-290	&	397	&	-9.7	&	0	&	0	&	-1.8	&	351	&	4365&4399&$D^{\pm}D_{0}^{0}$\\
$2^{5}S_{2}$	&	$2^{++}$	&	449.0	&	-290	&	397	&	9.7	&	0	&	0	&	-1.8	&	332	&	4385& ...& ...\\
$3^{1}S_{0}$	&	$0^{++}$	&	758.0	&	-215	&	578	&	-12.5	&	0	&	0	&	-1.2	&	407	&	4672& ...& ...\\
$3^{3}S_{1}$	&	$1^{+-}$	&	758.0	&	-215	&	578	&	-6.5	&	0	&	0	&	-1.2	&	401	&	4679& ...& ...\\
$3^{5}S_{2}$	&	$2^{++}$	&	758.0	&	-215	&	578	&	6.5	&	0	&	0	&	-1.2	&	389	&	4691& ...& ...\\

$1^{1}P_{1}$	&	$1^{--}$	&	355.2	&	-256	&	323	&	-13.0	&	0	&	0	&	-2.2	&	301	&	4268& ...& ...\\
$1^{3}P_{0}$	&	$0^{-+}$	&	355.2	&	-256	&	323	&	-6.4	&	-48.4	&	-38	&	-2.4	&	380	&	4187&4200&$D^{\pm}D^{*}_{0}$\\
$1^{3}P_{1}$	&	$1^{-+}$	&	355.2	&	-256	&	323	&	-6.4	&	-24.5	&	19.2	&	-1.8	&	300	&	4269&4290&$D^{\pm}D_{1}^{0}$\\
$1^{3}P_{2}$	&	$2^{-+}$	&	355.2	&	-256	&	323	&	-6.4	&	24.3	&	-3.8	&	-2.0	&	274	&	4294&4329&$D^{\pm}D^{*}_{2}$\\
$1^{5}P_{1}$	&	$1^{--}$	&	355.2	&	-256	&	323	&	6.4	&	-72.1	&	-27.6	&	-1.8	&	380	&	4188&4289&$D^{\pm}D_{1}$\\
$1^{5}P_{2}$	&	$2^{--}$	&	355.2	&	-256	&	323	&	6.4	&	-24.2	&	27.4	&	-2.2	&	278	&	4290&4431& $D^{*} D_{1}^{0}$\\
$1^{5}P_{3}$	&	$3^{--}$	&	355.2	&	-256	&	323	&	6.4	&	48.3	&	-7.7	&	-2.3	&	240	&	4327&4471&$D^{*}D^{*}_{2}$\\
$2^{1}P_{1}$	&	$1^{--}$	&	674.4	&	-188&	514	&	-10.2	&	0	&	0	&	-1.3	&	360	&	4590& ...& ...\\
$2^{3}P_{0}$	&	$0^{-+}$	&	674.4	&	-188	&	514	&	-5.1	&	-41.5	&	-32.5	&	-1.3	&	428	&	4521& ...& ...\\
$2^{3}P_{1}$	&	$1^{-+}$	&	674.4	&	-188	&	514	&	-5.1	&	-20.4	&	16.2	&	-1.3	&	359	&	4591& ...& ...\\
$2^{3}P_{2}$	&	$2^{-+}$	&	674.4	&	-188	&	514	&	-5.1	&	20.4	&	-3.5	&	-1.4	&	337	&	4613& ...& ...\\
$2^{5}P_{1}$	&	$1^{--}$	&	674.4	&	-188	&	514	&	5.1	&	-62.3	&	-23.2	&	-1.3	&	430	&	4520& ...& ...\\
$2^{5}P_{2}$	&	$2^{--}$	&	674.4	&	-188	&	514	&	5.1	&	-20.5	&	22.3	&	-1.4	&	342	&	4608& ...& ...\\
$2^{5}P_{3}$	&	$3^{--}$	&	674.4	&	-188	&	514	&	5.1	&	41.2	&	-6.5	&	-1.3	&	309	&	4641& ...& ...\\
[1ex]
\hline
\end{tabular*}
}

\setcounter{table}{4}
\begin{table}
\label{tab:7}
\tbl{Comparison of the tetraquarks masses from the present work with others}{

\begin{tabular}{llllllllllllll}
\toprule
$cq\bar{c}\bar{q}$ State & $J^{PC}$  & Ours &\cite{exotic4}& \cite{prd}& \cite{de}&\cite{mn}&\cite{smpatel}&\cite{farhad} &\cite{nr} &\cite{mh}& \cite{MV} \\ \colrule
$1^{1}S_{0}$ &$0^{++}$ & 3811&3812&4076&3852&3641&3849&3842&4056&3792&3729  \\
$1^{3}S_{1}$&$1^{+-}$&3846&3890&4156&3890&4047&3822&...&4079&...&3833 \\
$1^{5}S_{2}$&$2^{++}$&3914&3968&4262&3968&...&3922&...&4118&...&3988 \\

$1^{1}P_{1}$&$1^{--}$&4268&4350&4582&...&4545&...&...&...&4262&...\\

$1^{3}P_{0}$&$0^{-+}$&4187&4304&...&...&4567&...&4207&...&...&...\\

$1^{3}P_{1}$&$1^{-+}$&4269&4345&...&...&...&...&...&...&...&...\\

$1^{3}P_{2}$&$2^{-+}$&4294&4367&4585&...&...&...&...&...&...&...\\

$1^{5}P_{1}$&$1^{--}$&4188&4277&...&...&4570&...&...&...&...&...\\

$1^{5}P_{2}$&$2^{--}$&4290&4379&...&...&...&...&...&...&...&...\\

$1^{5}P_{3}$&$3^{--}$&4327&4381&4591&...&...&...&...&...&...&...\\
\hline
%
\end{tabular}
}
\end{table}


While the relativistic correction effects lead to make a very minor changes in the masses of tetraquarks, taking into account the spin degrees of freedom may lead to a little reduction or increase in the masses of tetraquarks.
The masses of 1S-wave tetraqurk states ($cq\bar{c}\bar{q}$) obtained from data set II are 100 MeV below the two-meson thresholds, implying that these states may be accounted by the two meson thresholds stated in Table:4. Initial predictions for $cq\bar{c}\bar{q}$ below the $D^{0}\bar{D}^{*0}$ threshold were made in Refs. \cite{de,de1} which was expected to be X(3872), and these were confirmed by subsequent works \cite{LM,LM1}.
The masses of 1S-wave of the $cq\bar{c}\bar{q}$ tetraquark state have been predicted by Carlucci et al., \cite{MV} to be 3.729-3.857 GeV, while the masses of the $bq\bar{b}\bar{q}$ tetraquark state have been expected to be a 10.260 and a 10.264 GeV for the $S\bar{S}$ (scalar-scalar) and $A\bar{A}$ (axial-axial) diquark–antidiquark, respectively. 
\\
Maiani et al. \cite{LM}, have also reported the masses of the 1S-wave of the $cq\bar{c}\bar{q}$ tetraquark as 3.723-3.832 GeV for the $S\bar{S}$ and $A\bar{A}$ diquark–antidiquark  configuration, respectively. In Ref.\cite{mh}, author predicted the Y(4260) state as 1P-wave and Y(4660) state as 2P-wave in $S\bar{S}$ system, while Y(4360) state as 1P-wave and Z(4430) state as 2S-wave in $A\bar{A}$ configuration. In the present study as well the mass range of 1P-wave state is found to 4.187-4.327 GeV. Particularly the mass of $1^3P_{1}$ state is 4.269 GeV which is close to the mass of Y(4260) and also to $D^{\pm}D_{1}^{0}$ meson threshold. The mass of $2^3P_{2}$ state which is close to experimentally observed state $\psi(4660)$. In Ref. \cite{prd}, also the author predicts the masses of $\psi(4660)$ state in 1P-wave with 100 MeV mass uncertainty and for $Z_{c}(3900)$, X(3915), $\psi(4360)$ states with 250 MeV mass uncertainties from experimental masses.
\\
\\
In the present study the mass of $1^3S_{1}$ state having quantum number $1^{+-}$ is 3846 MeV which is close to $Z_{c}(3900)$ and can exisit as $D^{\pm}D_{*}^{\pm}$ meson threshold. The mass of $Z_{c}(4430)$ could be explained as first radial excitation i.e. $2^3S_{1}$ of having quantum number $1^{+-}$ and can exist as a $D^{\pm}D_{0}^{0}$ meson threshold. There are experimental error bars in the mass predictions for these states. X(3940) with unmeasured quantum numbers might be a tetraquark state having quantum number ($2^{++}$) in $D_{*}^{\pm}D_{*}^{\pm}$ meson threshold of the $S\bar{S}$ tetraquark based on its mass value . The masses of charged $Z_{c}(4020)$, $Z_{c}(4050)$, $Z_{c}(4055)$, $Z_{c}(4100)$, and $Z_{c}(4200)$ are inconsistent with our findings. They might be hadro-charmonium or molecular states, for example \cite{exotic4}. The vector states Y(4260), and Y(4360) correspond to 1P-wave tetraquark states made up of $S\bar{S}$ and $A\bar{A}$ diquarks, respectively, whereas Y(4660) corresponds to the $S\bar{S}$ tetraquark's 2P-wave state. For the Y(4390) state, there is no tetra quark candidate from our model. Indeed, the masses of these states are projected to be within experimental error ranges. As we have shown in our model, the masses of $cq\bar{c}\bar{q}$ tetraquarks are sensitive to the parameters utilised in the effective potential, which suggests that the skew fit of the model to data set II may be a viable reason for this divergence. The mass of the tetraquark can be slightly higher than the sum of the masses of the two open charm singlets, because strong attraction in color singlet channels is stronger than in color anti-triplet channels. Thus, it is not surprising that the observed tetraquarks appear near to the corresponding meson thresholds, albeit being heavier.

\section{Summary}
In the current study, we have estimated the mass spectra of light-heavy tetraquarks $[cq][\bar{c}\bar{q}]$ in a non-relativistic framework that includes the cornell like potential as well as the relativistic correction term to the potential. The spin-dependent interactions have been introduced into the central potential in a perturbative manner in order to examine the splitting between distinct radial as well as orbital excitations. 
\\
\\
Tetraquarks have been hypothesised to be composed of axial-vector diquarks and antidiquarks in a colour antitriplet-triplet ($\bar{3}_{c}-3_{c}$) configuration being the most likely. We began by estimating the masses of charm mesons to fit the model's free parameters, and then computed the masses of scalar diquarks to get the mass spectra of corresponding tetraquarks without breaking the Pauli exclusion principle. In this method, we can anticipate the masses of diquarks and tetraquarks that contain charm and light quarks in their substructure. The double hidden-charm tetraquark states have a much higher energy than conventional charmonium mesons, and they can be distinguished experimentally from ordinary $q\bar{q}$ states by the presence of a hidden charm in their constituent.
\\
\\
We explored the most discussed $\chi_{1}(3872)$, $Z_{c}(3900)$, $Z_{c}(4430)$, Y(4260), Y(4360), $\psi(4660)$, states and compared them from present study.
The two charged $Z'_{c}$ states namely $Z_{c}(3900)$ and $Z_{c}(4430)$ having quantum number ($1^{+-}$) could be identified tetraquark state in which $Z_{c}(3900)$ belongs to ground state whereas $Z_{c}(4430)$ belongs to first radial excitation. Another state namely X(3940) whose quantum number is unmeasured yet and could be identified as ($2^{++}$). The vector states Y(4260) could be identified as 1P-wave state whereas Y(4360) which is a combination of $A\bar{A}$ tetraquark whose mass is nearly 100 MeV above from present study. The mass of the $\psi(4660)$ state matches with  $2^3 P_{2}$ state with quantum number $2^{-+}$ which is nearly 50 MeV below from the experimental mass.
\\
In the current spin-independent formalism, the theoretical uncertainties of our numerical findings for tetraquark masses are caused by the interaction of the uncertainties associated with the di-quark mass and the uncertainty associated with the potential parameters. The uncertainties included within the model may be assessed, and the majority of them are due to the approximations used. It is based on the examination of meson  mass spectra that the model's parameters, including quark masses and parameters of the interquark potential, are very firmly fixed in place.


%
%

\section*{Acknowledgments}
The authors are thankful to the organizers of 10th International Conference on New Frontiers in Physics (ICNFP 2021) for giving the opportunity to present our work.



\begin{thebibliography}{0}    

\bibitem{Y. R. Liu et al} Y. R. Liu et al., {\it Prog. Part. Nucl. Phys.}, \textbf{107}, 237–320 (2019).
\bibitem{S. L. Olsen} S. L. Olsen, T. Skwarnicki, D. Zieminska,  {\it Rev. Mod. Phys.}, \textbf{90}, 015003 (2018).

\bibitem{X3872} S.K. Choi et al., (LEPS Collaboration), {\it Phys. Rev. Lett.}, \textbf{91}, 262001 (2003).
\bibitem{cccc} R. Aaij et al., (LHCb Collaboration), {\it Scib}, \textbf{65}, 1983 (2020).
\bibitem {Y4260} B. Aubert et al. (BABAR Collaboration), {\it Phys. Rev. Lett.}, \textbf{95}, 142001 (2005).
\bibitem{Z3900} Z. Q. Liu et al. (Belle Collaboration),  {\it Phys. Rev. Lett.}, \textbf{110}, 252002 (2013).
\bibitem{exotic1} N. Brambilla et al., {\it Eur. Phys. J. C}, \textbf{71}, 1534 (2011).
\bibitem{exotic2} A. Esposito, A. Pilloni and A. D. Polosa, {\it Phys. Rept.}, \textbf{668}, 1 (2017).
\bibitem{exotic3} H. X. Chen et al., {\it Phys. Rept.}, \textbf{639}, 1 (2016).
\bibitem{exotic4} R. N. Faustov, V. O. Galkin, E. M. Savchenko, {\it Universe},  \textbf{7}, 94 (2021).
\bibitem {QQqq1} G. Yang , J. Ping and J. Segovia {\it Phys. Rev. D}, \textbf{101}, 014001 (2020).
\bibitem {QQqq2} E. Hernádez et al., {\it Phys. Lett. B}, \textbf{800}, 135073 (2020).
\bibitem{X3872BABAR} B. Aubert et al. (BABAR Collaboration), {\it Phys. Rev. D}, \textbf{71}, 071103 (2005).
\bibitem{X3872CDFII} D. Acosta et al., (CDF II collaboration) {\it Phys. Rev. Lett.}, 93, 072001 (2004).
\bibitem{X3872D0} V.M. Abazov et al., (D0 collaboration), {\it Phys. Rev. Lett.}, 93, 162002 (2004).
\bibitem{X3872LHCB} R. Aaij, et al., (LHCb Collaboration), {\it Eur. Phys. J. C} \textbf{72}, 1972 (2012).
\bibitem{X3872CMS} S. Chatrchyan, et al., (CMS Collaboration), {\it J. High Energy Phys.}, \textbf{04}, 154 (2013).
\bibitem{Ahmed Ali} A. Ali, J. Sören Lange , S. Stone, {\it Prog. Part. Nucl. Phys.}, \textbf{97}, 123–198 (2017).
\bibitem{H. J. Lipkin} H. J. Lipkin, {\it Phys. Lett. B}, \textbf{172}, 242 (1986).
\bibitem{J. P. Ader} J. P. Ader, J.M. Richard, and P. Taxil, {\it Phys. Rev. D}, \textbf{25}, 2370 (1982).
\bibitem{M. Karliner} M. Karliner and J. L. Rosner, {\it Phys. Rev. Lett.}, \textbf{119}, 202001 (2017).
\bibitem{E. J. Eichten} E. J. Eichten and C. Quigg, Phys. Rev. Lett. \textbf{119}, 202002 (2017).
\bibitem{A. Czarnecki} A. Czarnecki, B. Leng, and M. B. Voloshin, {\it Phys. Lett. B},
\textbf{778}, 233 (2018).
 
\bibitem{S. Q. Luo} S. Q. Luo et al., {\it Eur. Phys. J. C}, \textbf{77}, 709 (2017).

\bibitem{sasa} S. Prelovsek, Proceedings, 32nd International Symposium on Lattice Field Theory (Lattice 2014) PoS LATTICE2014 (2014) 015. arXiv:1411.0405.
\bibitem{exotic5} F.-K. Guo et al., {\it Rev. Mod. Phys.}, \textbf{90}, 015004 (2018).
\bibitem{E. Eichten1978} E. Eichten et. al, {\it Phys. Rev. D}, \textbf{17}, 3090 (1978).
\bibitem{E. Eichten1980} E. Eichten et.al, {\it Phys. Rev. D}, \textbf{21}, 203 (1980).
\bibitem{N. Brambilla} N. Brambilla, et al., {\it Eur. Phys. J. C}, \textbf{71}, 1534 (2011).
\bibitem{A. Gray} A. Gray et.al, {\it Phys. Rev. D}, \textbf{72} 094507 (2005).
\bibitem{R.J. Dowdall} R. J. Dowdall, et al., (HPQCD Collaboration), {\it Phys. Rev. D}, \textbf{85}, 054509 (2012).
\bibitem{L. Liu} L. Liu et.al, (Hadron Spectrum Collaboration), {\it J. High
Energy Phys.}, \textbf{07} 126 (2012).
\bibitem{Gyang} G. Yang, J. Ping, and J. Segovia, {\it Symmetry}, \textbf{12}, 1869 (2020).
\bibitem{L. Leskovec} L. Leskovec et al., {\it Phys. Rev. D}, \textbf{100}, 014503 (2019).

\bibitem{prd} P. Lundhammar and T. Ohlsson, {\it Phys. Rev. D}, \textbf{102}, 054018 (2020).

\bibitem{relativistic} J. M. Richard, A. Valcarce, and J. Vijande, {\it Phys. Rev. D}, \textbf{103}, 054020 (2021).

\bibitem{Pan} P. P. Shi, F. Huang, and W. L. Wang, {\it Phys. Rev. D}, \textbf{103}, 094038 (2021).
\bibitem{d1} M. Abud, F. Buccella, and F. Tramontano, {\it Phys. Rev. D}, \textbf{81}, 074018 (2010).
\bibitem{d2} R. F. Lebed and A. D. Polosa, {\it Phys. Rev. D}, \textbf{93}, 094024 (2016).
\bibitem{d3} J.Wu et. al {\it Phys. Rev. D}, \textbf{94}, 094031 (2016).
\bibitem{d4} R. F. Lebed, {\it Phys. Rev. D}, \textbf{96}, 116003 (2017).
\bibitem{d5} A. Ali et al., {\it Eur. Phys. J. C}, \textbf{78}, 29 (2018).
\bibitem{d6} J. F. Giron and R. F. Lebed, {\it Phys. Rev. D}, \textbf{101}, 074032 (2020).
\bibitem{d7} J. F. Giron, R. F. Lebed, and C. T. Peterson, {\it J. High Energy Phys.}, \textbf{01} 124 (2020).
\bibitem{d8} J. F. Giron and R. F. Lebed, {\it Phys. Rev. D}, \textbf{102}, 014036 (2020).
\bibitem{d9} P. P. Shi, F. Huang, and W. L. Wang, {\it Eur. Phys. J. C}, \textbf{79}, 314 (2019).
\bibitem{rohit} R. Tiwari, D. P. Rathaud, and A. K. Rai, {\it Eur. Phys. J. A}, \textbf{57}, 289 (2021). 
\bibitem{spin2} W. Lucha, F. F. $Sch\ddot{o}berl$, and D. Gromes, {\it Phys. Rept.,} \textbf{200}, 127 (1991).
\bibitem{scr1} S. Godfrey and N. Isgur, {\it Phys. Rev. D}, \textbf{32}, 189 (1985).
\bibitem{scr2} D. M. Brink and F. Stancu, {\it Phys. Rev. D}, \textbf{49}, 4665 (1994).
\bibitem{cornell} E. Eichten et al., {\it Phys. Rev. D} \textbf{21}, 203 (1980).
\bibitem{koma} Y. Koma, M. Koma, H. Wittig, {\it Phys. Rev. Lett.}, \textbf{97}, 122003 (2006).
\bibitem{dpepjc}  A. K. Rai and D. P. Rathaud, {\it Eur. Phys. J. C}, \textbf{75}, 462 (2015).

\bibitem{ak} A. K. Rai, J. N. Pandya and P. C. Vinodkumar, {\it Nucl. Phys. A}, {\bf{782}}, 406 (2007).
\bibitem{vk} V. Kher, A. K. Rai {\it Chin. Phys. C},  \textbf{42}, 083101 (2018).
\bibitem{dpepjp} D. P. Rathaud and A. K. Rai {\it Eur. Phys. J. Plus}, \textbf{132}, 370 (2017).
\bibitem{dpijp1} D. P. Rathaud and A. K. Rai {\it Indian J. Phys}, \textbf{90}, 1299 (2016).
\bibitem{dpijp2} D. P. Rathaud and A. K. Rai {\it Indian J. Phys}, {\bf 95}, 2807 (2021).

\bibitem{dpfew} D. P. Rathaud and A. K. Rai, {\it Few-Body Syst}, \textbf{60}, 1 (2019).
\bibitem{spin1} M. B. Voloshin, Prog. {\it Part. Nucl. Phys.}, \text{61}, 455 (2008).

\bibitem{griffiths} D. Griffiths, {\it Int. to Elementary Particles, Second Revised Edition, Wiley-VCH} (2008).
\bibitem{debastiani} V. Debastiani and F. Navarra, Chin. Phys. C \textbf{43}, 013105 (2019).
\bibitem{bethe} H. A. Bethe and E. E. Salpether, {\it Quantum Mechanics of atoms of one- and two-electrons, Springer} (1957).
\bibitem{thesis} V. R. Debastiani, {\it Spectroscopy of the All-Charm Tetraquark, Master thesis} (2016)
\bibitem{tb} C. Cohen-Tannoudji, B. Diu, and F. Laloe, {\it Quantum Mechanics, Vol. 2, Wiley-VHC} (1978).
\bibitem{pdg} P. Zyla et al., (Particle Data Group), {\it Prog. Theor. Exp. Phys.}, 083C01 (2020).
\bibitem {de} D. Ebert, R. Faustov, and V. Galkin, {\it Phys. Lett. B} \textbf{634}, 214 (2006).
\bibitem {de1} D. Ebert, R. N. Faustov, and
V. O. Galkin, {\it Phys. At. Nucl.}, \textbf{72}, 184 (2009).

\bibitem {LM} L. Maiani, F. Piccinini, A. Polosa, and V.
 Riquer, {\it Phys. Rev. D} 71, 014028 (2005).
\bibitem{RK} R. Kleiv et al., {\it Phys. Rev. D} \textbf{87}, 125018 (2013).
\bibitem {LM1} L. Maiani et al., {\it Phys. Rev. Lett.} 93, 212002 (2004)
\bibitem {mn} M. N. Anwar, J. Ferretti, and E. Santopinto, {\it Phys. Rev. D} \textbf{98}, 094015 (2018).
\bibitem{smpatel} S. Patel, P.C. Vinodkumar, {\it Eur. Phys. J. A}, \textbf{50}, 131 (2014).



\bibitem{farhad} F. Zolfagharpour and M. Aslanzadeh, {\it Eur. Phys. J. A},  \textbf{55}, 86 (2019).
\bibitem{nr} D. Ebert et al., {\it Phys. Rev. D} \textbf{76}, 114015 (2007). 
\bibitem{mh} M. Hadizadeh and A. Khaledi-Nasab, {\it Phys. Lett. B}, \textbf{753}, 8 (2016).
\bibitem{MV} M. V. Carlucci1 et al., {\it Eur. Phys. J. C},
57, 569 (2008).

































































%

%

%
%
%
%

%
%
\end{thebibliography}
\end{document}